\documentclass[8.5pt,twoside,twocolumn]{article}
\usepackage[super,sort&compress,comma]{natbib} 
\usepackage{mhchem}
\usepackage{times,mathptmx}
\usepackage{sectsty}
\usepackage{balance} 
\usepackage{esint}
\usepackage{hyperref}

\usepackage{graphicx} 
\usepackage{lastpage}
\usepackage[format=plain,justification=raggedright,singlelinecheck=false,font=small,labelfont=bf,labelsep=space]{caption} 
\usepackage{fancyhdr}
\begin{document}

\twocolumn[
  \begin{@twocolumnfalse}
\noindent\LARGE{\textbf{Reversal and tunability of the surface plasmon enhanced optical forces on a nanorod pair in the presence of a dielectric interlayer$^\dag$}}
\vspace{0.6cm}

\noindent\large{\textbf{Aybike Ural Yal$\c{c}\i$n,\textit{$^{a}$} \"{O}zg\"{u}r E. M\"{u}stecapl{\i}o\u{g}lu,\textit{$^{a}$} and
Kaan G\"{u}ven$^{\ast}$\textit{$^{a}$}}}\vspace{0.5cm}

\noindent \normalsize{We investigate numerically the modification of the surface-plasmon enhanced optical force on a gold nanorod pair by incorporating a dielectric interlayer. The frequency dependent electromagnetic forces are obtained through the full-vectorial solution of the Maxwell's equations by a finite element solver. We obtain the common and the relative electromagnetic force experienced by the nanorod pair in the presence of a dielectric interlayer of different permittivity or thickness. In particular, we demonstrate that a liquid crystal dielectric interlayer can be utilized as a dynamic tuning mechanism for the magnitude and the reversal of the direction of the relative force.}
\vspace{0.5cm}
 \end{@twocolumnfalse} 
  ]

\section{Introduction}


\footnotetext{\textit{$^{a}$Department of Physics, Ko$\c{c}$ Univers$\i$ty, Sar{\i}yer, Istanbul 34450, Turkey}}

The utilization of optical forces to control the position of micro- to nano-scale particles is advancing rapidly with emerging chemical, biomedical and integrated photonics applications \cite {van_thourhout_optomechanical_2010,li_harnessing_2008,simmons_quantitative_1996,svoboda_biological_1994}.  Optical forces can be generally categorized as gradient- and scattering-type forces. The gradient force is generated by the refraction or diffraction of the optical beam by the target particle, forming the basic mechanism of optical tweezers \cite {svoboda_biological_1994}. The scattering force is an axial force on the particle due to the momentum of the impinging beam and it  is used in cavity optomechanics \cite{kippenberg_cavity_2008}.

These forces are widely used for micron-sized particles (approx. few wavelengths in the visible spectrum). For a significantly measurable amount of optical force in the subwavelength regime, however, the enhancement of the local electromagnetic field by the surface-plasmons (SPs) on the target particle  becomes essential. For instance, in a nanoparticle system consisting of two closely placed metallic spheres, when the frequency of the illuminating laser is close to the SP resonance of the metallic spheres, a strong enhancement in the optical force can be obtained \cite{chu_laser-induced_2007,halterman_plasmonic_2005,arias-gonzalez_optical_2003,hallock_optical_2005}.

In this study, we investigate tailoring the magnitude and the direction of optically induced forces on metallic nanoparticles via geometric and material parameters of a dielectric interlayer, which can be utilized for nanoscale applications. In particular, we propose that a tunable force mechanism between the nanoparticles is possible by inserting a liquid crystal layer as in ref.~\cite{zografopoulos_liquid-crystal-tunable_2013}.

The paper is organized as follows:  Section 2 gives a brief theory on force calculation through Maxwell Stress Tensor. In Section 3, we describe the nanorod-pair model for investigating the SP-assisted forces and the simulation method. Section 4 presents the results obtained with a dielectric layer of varying dielectric constant and thickness. The tunability of the optical force via liquid crystal dielectric layer is discussed in Sec.~5.  Section 6 summarizes the major outcomes of the study and concludes the paper.

\section{Methods and Theory}

\begin{figure}[h]
\centering
\includegraphics[width=8.3cm,height=6.3cm]{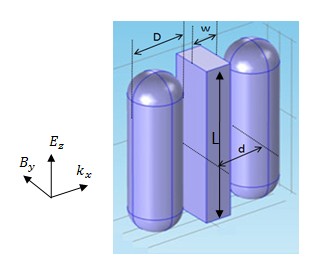}
\caption{The isometric view of the nanorod-pair-dielectric interlayer model. The geometrical parameters of the model and the incident electromagnetic field are indicated.}
\label{fig1}
\end{figure}

We adopt the nanorod-pair model discussed in a recent study \cite{zhao_optical_2010} in order to relate our results with the existing literature. This model consists of two identical pill-shaped metal nanorods, each with a cylindrical body and hemispherical caps on either end (see Fig.~\ref {fig1}). The nanorods are of length L and diameter D. The distance between the center axes of the rods is d + D. The dielectric interlayer is a rectangular slab of length $L$, width $D$, and thickness $w$. The coordinate system is set such that the rods have their center on the x-axis, extend in the z-direction longitudinally and in the y-direction laterally. The electromagnetic wave is linearly polarized in the z-direction and incident in the x-direction to the nanorods.

We performed the numerical simulations using a commercial software (Comsol Multiphysics) which employs a finite element solver for Maxwell equations. The 3D model system is placed at the center of a large spherical computation domain, which is encapsulated by perfectly matched layers to simulate open space (reflectionless) boundary conditions. The nanorod material is chosen as gold with frequency dependent permittivity \cite{johnson_optical_1972}. In this work, we consider a lossless dielectric. The incident electromagnetic (em) radiation wavelength ranges from $600$ nm to $800$ nm.

Once the convergence of the electromagnetic field solutions is obtained at a given wavelength through the simulation, the optical force on each nanorod is calculated using the Maxwell stress tensor. The $i^{th}$ component of the total Lorentz force is given by
 \begin{align}
    F_i=\iiint\limits_V (\rho E_i+\epsilon_{ijk}J_jB_k) dV
  \end{align}
where  $\epsilon_{ijk}$ is Levi-Civita symbol and V  is a volume that encloses a nanorod. Defining force per unit volume expressed in terms of divergence of Maxwell stress tensor $\overleftrightarrow{T}$ and partial time derivative of Poynting’s vector $\overrightarrow{S}$, the integral can be transformed to
\begin{align}
   F_i=-\varepsilon_0 \mu_0 \iiint\limits_V \frac{\partial S_i}{\partial t} dV + \iiint\limits_V \frac{\partial T_{ij}}{\partial x_j} dV
  \end{align}
where $S_i=\varepsilon_0 c^2 \epsilon_{ijk} E_j B_k$ is Poynting’s vector($c=\text{speed of light in free space}$ and $\varepsilon_0:\text{vacuum permittivity}$) and
\begin{align}
    T_{ij}= \varepsilon_0 ( E_i E_j -\frac{1}{2}\delta_{ij} E^2  ) +\frac{1}{\mu_0} ( B_i B_j -\frac{1}{2}\delta_{ij} B^2 )
 \end{align}
with $\delta_{ij}$ Kronecker delta. In the steady state, the time average of the first term on the right-hand side of Eq. (2) vanishes. The second term can be converted to a flux integral 
\begin{align}
   \langle F_i \rangle=\oiint\limits_S \langle T_{ij}\rangle n_jdS
 \end{align}
over an integration surface S that encapsulates each nanorod in separate where $n_j$ is the outward surface normal in the $j^{th}$ direction.

In order to better understand the behavior of the optical force depending on the frequency of the incident field, the electric and magnetic dipole moments of the nanorod pair are calculated. The electric dipole moment is given by
\begin{align}
   p_z=\iiint_V P_z dV= \iiint_V (D_z-\varepsilon_0 E_z)dV
 \end{align}
where $P_z$ is the z-component of the polarization and $D_z$ is the z-component of the displacement field.
The magnetic moment is calculated from the rotation of the polarization currents $r\times J_p$ (r is the position vector pointing the volume element that contains the current density from the origin which is the center of mass (CM) of the nanorod pair ) , which yields\begin{align}
   m_y=\iiint_V (z\frac{\partial P_x}{\partial t}-x\frac{\partial P_z}{\partial t}) dV\\
          =-i\omega \iiint_V [z(D_x-\varepsilon_0 E_x)-x(D_z-\varepsilon_0 E_z)]dV
 \end{align}
where $\omega$ is the frequency of the time harmonic fields. 

\section{Optical force in a gold nanorod pair system}

We first calculate a reference force spectra for the gold nanorod pair in vacuum, as in Ref\cite{zhao_optical_2010}. We take the gold nanorods’ diameter $D=25$ nm, length $L=100$ nm and the distance between the rod's axes $d=35$ nm. Henceforth, we refer to the $x$-component of the force without the coordinate subscript. We label the nanorods as the $2^{nd}$ and the $1^{st}$ in the direction of incidence of the electromagnetic wave (i.e. along the propagation direction of the incident em wave). The common force $F_{common}=(F_1+F_2 )$ is the force acting on the center of mass of the nanorods.The relative force is defined as $F_{relative}=(F_1 -F_2)/2$ . According to this definition, a positive relative force is repulsive whereas the negative relative force is attractive.

\begin{figure}[h!]
\centering
\includegraphics[width=8.3cm]{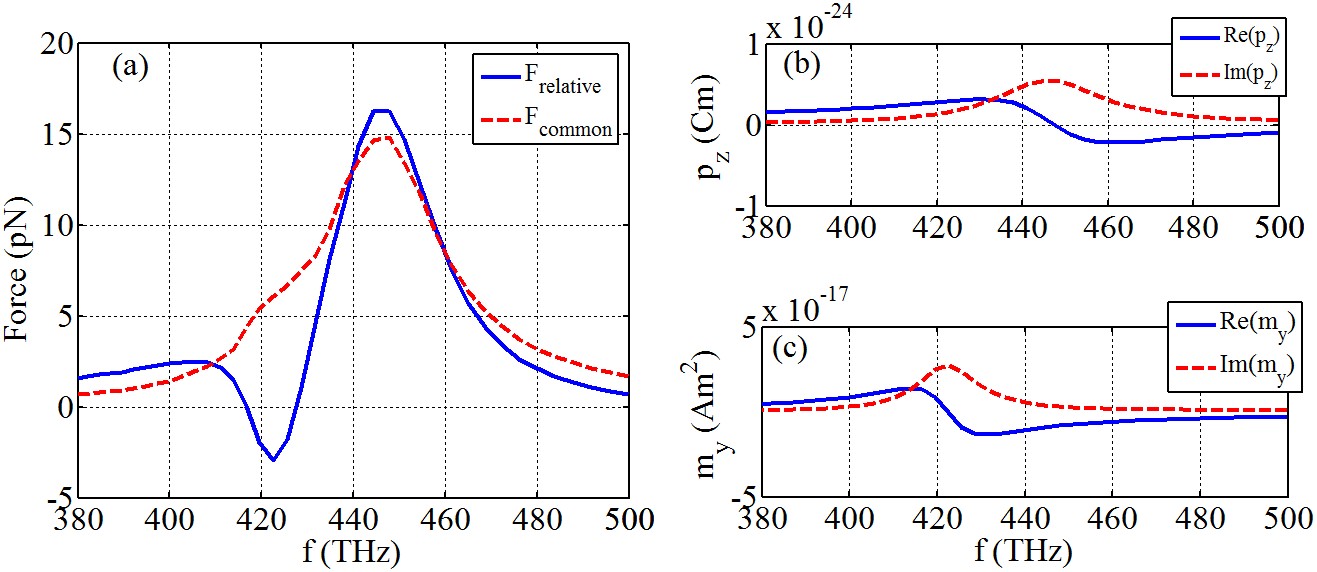}
\caption{(a) The frequency spectra of the relative (solid blue curve) and common (dashed red curve) force in the nanorod pair system. The nanorods are of length L=100 nm, diameter D=25 nm and inter-rod distance d=35 nm. (b-c) The real (solid blue curve) and imaginary (dashed red curve) parts of the (b) electric and (c)  magnetic dipole moment of the nanorod pair.}
\label{fig2vacuumfcomfreldipoles}
\end{figure}

Figure \ref {fig2vacuumfcomfreldipoles}(a) shows the calculated common- and relative force spectra of the reference system. The common force is the axial force on the center of mass of the nanorod pair, hence it is positive. Both the common and the relative force exhibit peaks at 422 THz and 445 THz. Note that the relative force takes negative and positive values respectively at the peak frequencies. The electric and magnetic dipole moments plotted in Fig.~\ref {fig2vacuumfcomfreldipoles}(b)-(c) indicate that the peaks in the force spectrum are associated with the dipole excitations of the nanorod pair. The antisymmetric eigenmode at 422 THz corresponds to the magnetic dipole moment and leads to an attractive relative force, whereas the symmetric eigenmode at 445 THz is the electric dipole moment and results in the repulsive relative force. 

The optical force depends on the geometrical parameters of the nanorod pair, in particular to the rod length and the inter-rod distance but the spectral features remain similar. We refer the reader to Ref.\cite{zhao_optical_2010} for details of this analysis.

\section{Optical force between gold nanorods in the presence of a dielectric interlayer}

We now discuss the effect of the dielectric interlayer on the optical force acting on the nanorods. Throughout this analysis, the geometrical parameters of the reference model are preserved. As depicted in Fig.~\ref{fig1}, the dielectric layer has a constant length $L=100$ nm and the width is equal to $D=25$ nm. We assume that the dispersion of the dielectric material is constant and the dissipation is negligible, which may be justified for a nanometer thick dielectric layer subject to THz frequency fields. We characterize the optical force for the different thickness (w) and the dielectric permittivity ($\varepsilon$) of the dielectric interlayer.

\begin{figure}[h!]
\centering
\includegraphics[width=8.3cm]{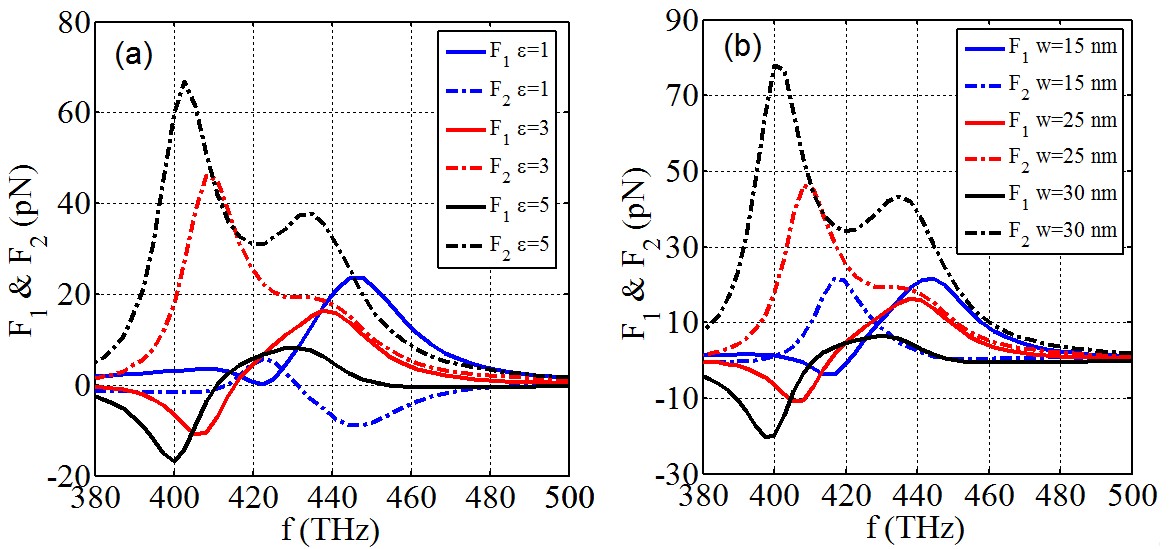}
\caption{The calculated individual force spectra of nanorods in the presence of a dielectric interlayer for (a) fixed w = 25 nm, $\varepsilon= 1, 3, 5$ and for (b) fixed  $\varepsilon= 3$ w = 15, 25, 30 nm, respectively.}
\label{fig3}
\end{figure} 
Figure \ref {fig3} shows the optical force acting on each nanorod individually, in panel (a) for fixed dielectric layer thickness of $w = 25$ nm with different permittivity values, and in panel (b) for fixed permittivity of $\varepsilon= 3$ with different thicknesses. Evidently, increasing the electromagnetic field intensity between the rods (either by increasing the permittivity or the volume of the dielectric material) enhances the electromagnetic force. Increasing the permittivity or the thickness induces a red shift of the force peaks that are associated with the electric and magnetic dipole resonances. In Figure \ref {fig3} (a), the $F_1$ spectra (solid curves) show a rather uniform negative shift, whereas the $F_2$ spectra (dashed curves) shift rapidly and nonuniformly in the positive direction with increasing permittivity. We have found that the electric dipole resonance peak of $ F_2$ changes from negative to positive for $\varepsilon> 1.7$. Figure \ref {fig3} (b) shows that, increasing the thickness of the dielectric layer induces similar changes in the force spectra.

\begin{figure}[h!]
\centering
\includegraphics[width=8.3cm]{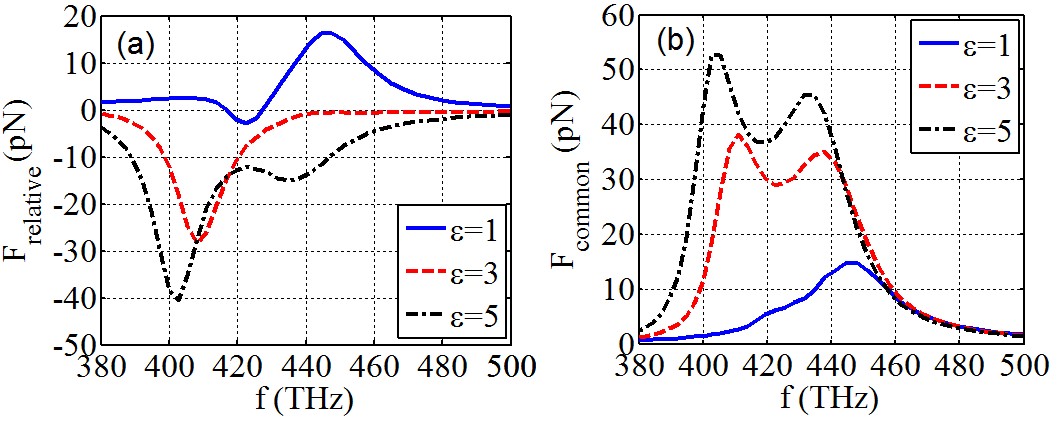}
\caption{The (a) relative and (b) common force spectra of nanorod pair with a dielectric interlayer of  different permittivities and of a constant thickness $w= 25$ nm.}
\label{fig4}
\end{figure}

As a consequence, the relative force spectra changes significantly as shown in Fig.~\ref {fig4} (a). In particular, the relative force at the electric dipole resonance can be repulsive (for $\varepsilon = 1$), negligibly small (for $\varepsilon = 3$), or attractive (for $\varepsilon =5$). Therefore, the direction of the relative force may not be deduced solely from the symmetry of the excitation mode of the nanorods as it was in the absence of the dielectric layer. The polarization of the dielectric layer contribute to the local field distribution at the nanorod surfaces where the optical forces are calculated. The relative force at the magnetic dipole resonance remains attractive and increases in magnitude. The common force plotted in Fig.~\ref {fig4} (b) is always repulsive, as expected.

\begin{figure}[h!]
\centering
\includegraphics[width=8.3cm]{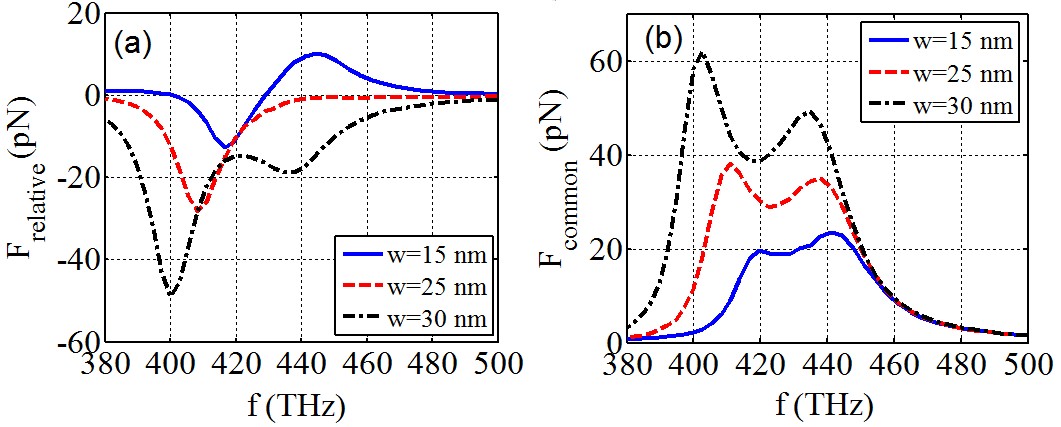}
\caption{The (a) relative and (b) common force spectra of nanorod pair with a dielectric interlayer of  different thicknesses and of a constant permittivity $\varepsilon=3$.}
\label{fig5}
\end{figure}

In Figure \ref {fig5}, we plot the relative and common force for different dielectric layer thicknesses at fixed permittivity $\varepsilon=3$. Again, the relative force  at the electric dipole resonance frequency changes sign with increasing dielectric layer thickness.

\begin{figure}[h!]
\centering
\includegraphics[width=8.3cm]{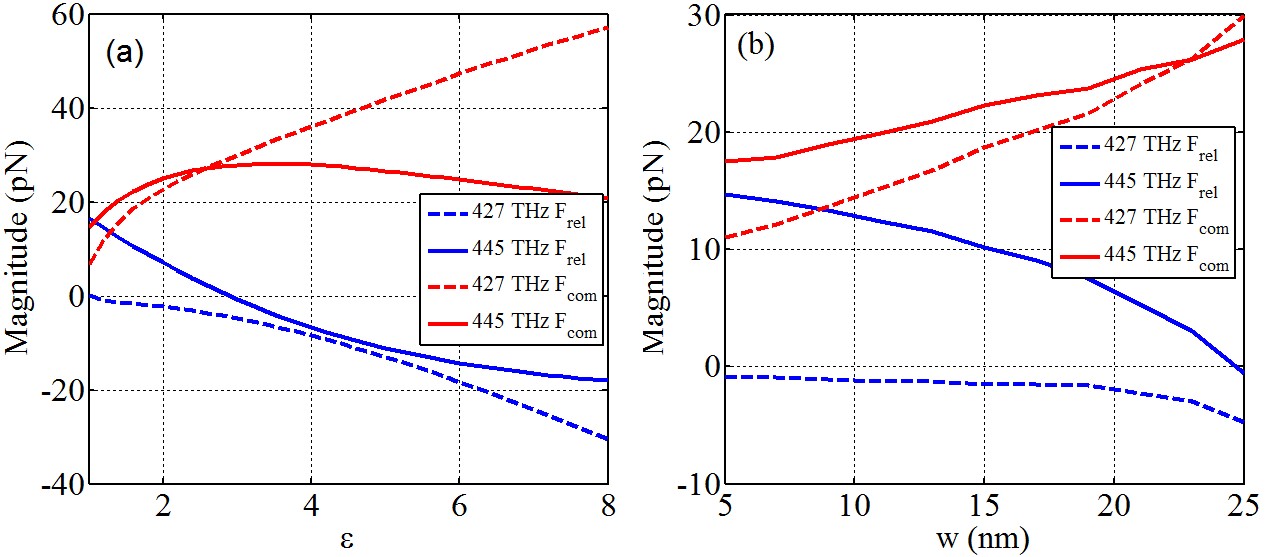}
\caption{The common- (upper two curves) and the relative force (lower two curves) as a function of permittivity for a 25 nm thick dielectric interlayer at 427 THz (dotted) and 445 THz (solid); (b) Same as in (a) but as a function of thickness of a dielectric interlayer of $\varepsilon= 3$.}
\label{fig6}
\end{figure}

In Figures \ref {fig6} (a) and (b), we plot the optical force at selected frequencies as a function of the permittivity with fixed thickness ($w = 25$ nm) and as a function of the thickness with fixed permittivity ($\varepsilon=3$), respectively.  In Figure \ref {fig6} (a), the relative force  is all attractive for 427 THz (dashed lower curve), whereas it exhibits a sign change for 445 THz at $\varepsilon= 3$ (solid lower curve). The common force at 427 THz (upper dotted curve) is increasing monotonically with increasing permittivity. The common force at 445 THz (upper solid curve) increases up to $\varepsilon= 3$, and decreases afterwards. In Figure \ref {fig6} (b), the relative force at 445 THz (lower solid curve) decreases monotonically and reverses sign at w=24.5 nm. The relative force at 427 THz (lower dotted curve) is attractive throughout the range. The common force at both frequencies (upper two curves) increases with increasing thickness.

\section{Tunable relative force by the use of liquid crystals}
\begin{figure}[h!]
\centering
\includegraphics[width=8.3cm]{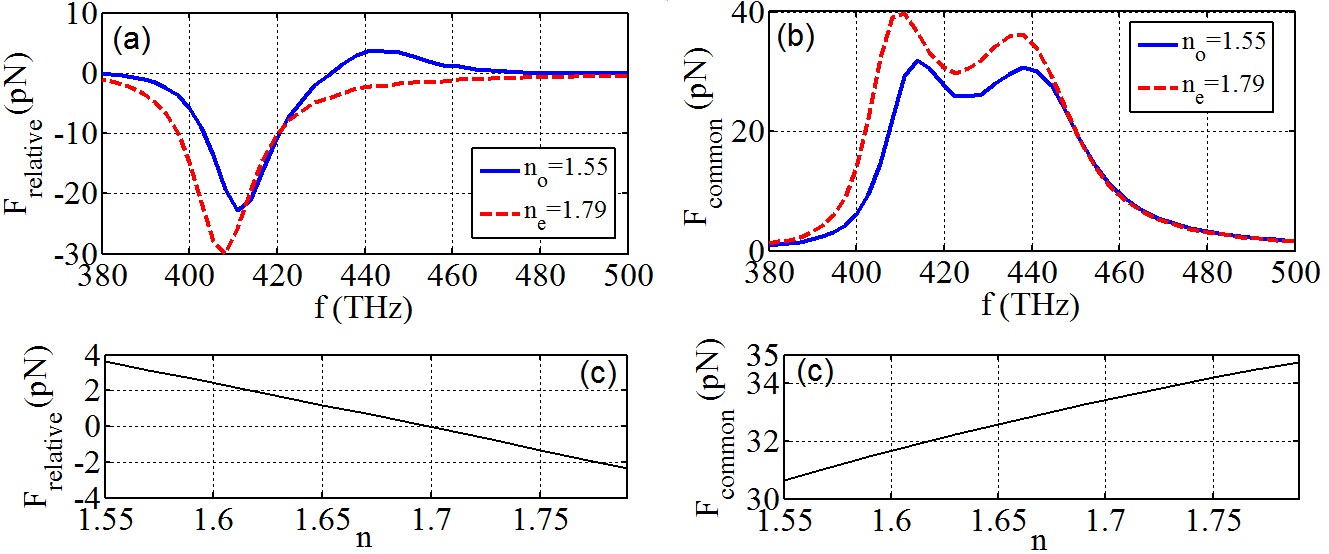}
\caption{The (a) relative and (b) common force spectra with a liquid crystal interlayer of thickness $w =25$ nm with the two extreme values of its refractive index from 1.55 to 1.79 under the voltage effect. (c) The relative (blue solid) and common force (red dotted) plotted as a function of refractive index at 440 THz.}
\label{fig7}
\end{figure}

 The aforementioned results suggest that a tunable optical force may be achieved by incorporating a dielectric material whose index of refraction can be controlled externally. Here, we propose the use of a birefringent liquid crystal whose index of refraction can be tuned between two values by an external electric field. In Figure \ref {fig7}, we plot the relative and common force in the presence of a liquid crystal interlayer for the ordinary and extraordinary indices of refraction (values taken from Ref.\cite{zhang_novel_2001}). For the nanorod pair model in this paper, a 25 nm thick dielectric interlayer can tune the relative optical force in an attractive/repulsive regime at an operation frequency of 440 THz, as shown in Fig.~\ref {fig7} (c).

\section{Conclusions}
In this study, we investigated the effect of a dielectric interlayer on the surface-plasmon enhanced optical forces acting on a nanorod pair, under electromagnetic excitation in THz regime. We found that the optical forces depend on the permittivity or the thickness of the dielectric. In particular, the relative force can be altered to become attractive or repulsive. This behavior can be utilized as a tunable optical force in plasmonic nanostructures, e.g. by employing liquid crystals.

\section{Acknowledgment}
{This work is supported by  the Science and Technology Research Council of Turkey (T\"UBITAK) under Project no. 111T285. K. G\"uven acknowledges partial support from the Turkish Academy of Sciences.}

\footnotesize{
\bibliography{rsc1} 
\bibliographystyle{unsrt} 
}

\end{document}